\documentclass[aps,twocolumn,showpacs,amsmath,amssymb]{revtex4}
\usepackage[dvips]{graphicx}
\usepackage[usenames]{color} 
\usepackage{bm}% bold math
\bmdefine{\bk}{k}
\bmdefine{\bv}{v}
\bmdefine{\bx}{x}
\bmdefine{\br}{r}
\bmdefine{\bp}{p}
\bmdefine{\bq}{q}

\newcommand{\rb}{\bar{\rho}}
\newcommand{\lka}{\left(}
\newcommand{\rka}{\right)}
\newcommand{\ltka}{\left\{}
\newcommand{\rtka}{\right\}}
\newcommand{\ldka}{\left[}
\newcommand{\rdka}{\right]}
\newcommand{\dbk}{\delta_\bk}
\newcommand{\dini}{\delta_{\rm ini}}
\newcommand{\ubk}{u_\bk}
\newcommand{\calH}{\mathcal{H}}
\newcommand{\calHk}{\mathcal{H}_k}
\newcommand{\calD}{\mathcal{D}}
\newcommand{\calDk}{\mathcal{D}_k}
\newcommand{\calP}{\mathcal{P}}
\begin{document}
\draft

%<<<<<<<<<<<<< TITLE >>>>>>>>>>>>>>>%
\title{Galaxy clustering constraints on deviations from Newtonian
gravity at cosmological scales II: Perturbative and numerical analyses
of power spectrum and bispectrum}

%<<<<<<<<<<<<< AUTHOR >>>>>>>>>>>>>>>%
\author{
Akihito Shirata$^{(1,2)}$,
Yasushi Suto$^{(2,3)}$,
Chiaki Hikage$^{(4)}$
\footnote{
Present address: School of Physics and Astronomy, University of Nottingham,
University Park, Nottingham NG7 2RD, United Kingdom}
,
Tetsuya Shiromizu$^{(1)}$,
Naoki Yoshida$^{(4)}$}

%<<<<<<<<<<<<< ADDRESS >>>>>>>>>>>>>>>%
\affiliation{$^{(1)}$Department of Physics, Tokyo Institute of Technology, 
Tokyo 152-8551, Japan}

\affiliation{$^{(2)}$Department of Physics, The University of Tokyo,  Tokyo
113-0033, Japan}

\affiliation{$^{(3)}$Research Center for the
Early Universe, The University of Tokyo,  Tokyo
113-0033, Japan}

\affiliation{$^{(4)}$Department of Physics, 
Nagoya University, Nagoya 464-8602, Japan}

%<<<<<<<<<<<<< DATE >>>>>>>>>>>>>>>%
\date{\today}

%======================================%
%<<<<<<<<<<<<< ABSTRACT >>>>>>>>>>>>>>>%
%======================================%
\begin{abstract}
We explore observational constraints on possible deviations from
Newtonian gravity by means of large-scale clustering of galaxies. We
measure the power spectrum and the bispectrum of Sloan Digital Sky
Survey galaxies and compare the result with predictions in an empirical
model of modified gravity.  Our model assumes an additional Yukawa-like
term with two parameters that characterize the amplitude and the length
scale of the modified gravity.  The model predictions are calculated
using two methods; the second-order perturbation theory and direct
$N$-body simulations.  These methods allow us to study non-linear
evolution of large-scale structure. Using the simulation results, we
find that perturbation theory provides reliable estimates for the power
spectrum and the bispectrum in the modified Newtonian model.  We also
construct mock galaxy catalogues from the simulations, and derive
constraints on the amplitude and the length scale of deviations from
Newtonian gravity. The resulting constraints from power spectrum are
consistent with those obtained in our earlier work, indicating the
validity of the previous empirical modeling of gravitational
nonlinearity in the modified Newtonian model. If linear biasing is
adopted, the bispectrum of the SDSS galaxies yields constraints very
similar to those from the power spectrum. If we allow for the nonlinear
biasing instead, we find that the ratio of the quadratic to linear
biasing coefficients, $b_2/b_1$, should satisfy $-0.4 < b_2/b_1<0.3$ in
the modified Newtonian model.
\end{abstract}


\maketitle

\label{sec:introduction}
\section{Introduction}

Recent measurements of cosmic microwave background radiation angular
power spectrum \cite{Spergel:2006hy, Bennett:2003bz, Spergel:2003cb}
strongly support the ``standard'' cosmological model in which the energy
content of the universe is dominated by dark energy (very close to
Einstein's cosmological constant, $\Lambda$) and cold dark matter
(CDM). Such $\Lambda$CDM universes are also in good agreement with
independent datasets of galaxy clustering \cite{Tegmark:2004uf,
Cole:2005sx} and distant Type Ia supernovae \cite{Knop:2003iy, Astier:2006pq,
Riess:2007rs}.  Thus the basic framework for the theory of structure
formation in the universe is firmly established. Nevertheless the nature
and the physical origin of dark energy remain to be understood.
 
The apparent accelerating expansion of the universe is conventionally
interpreted in terms of a source of repulsive force (dark energy), but
can be explained by modifying Newton's law of gravity on cosmological
scales as well.  The latter resolution has been seriously considered
recently.  For example, Dvali, Gabadadze and Porrati (DGP)
\cite{Dvali:2000hr,Deffayet:2000uy} propose that gravity leaking into
extra dimensions drives the observed accelerating expansion.  Other such
models include modified Newtonian dynamics (MOND) \cite{Sanders:2002pf,
Scarpa:2006cm, Bekenstein:2004ne} and ghost condensation
\cite{Arkani-Hamed:2003uy, Arkani-Hamed:2003uz}.  Intriguingly, {\it
all} of these alternative models predict some deviation from
conventional Newtonian gravity at cosmological scales.

Indeed, while the validity of Newtonian gravity is tested to high
precision up to the scale of the solar system ($\sim 10^{13}$ m), there
have been no rigorous tests at sub-millimeter and over scales beyond the
solar system \cite{Fischbach:1999bc,Adelberger:2003zx,Hoyle:2004cw}.  It
has been suggested that large-scale galaxy clustering can be used to
constrain non-Newtonian models of gravity \cite{Frieman:1991} in
principle, but it became feasible only recently with accurate
measurements of galaxy clustering in large redshift
surveys \cite{Tegmark:2004uf,Cole:2005sx}.

In our earlier work \cite{Shirata:2005yr} (Paper I), we put quantitative
constraints on deviations from Newtonian gravity at cosmological scales
under the assumption that the deviation can be described in a simple
parametric form; we adopted an empirical Yukawa-like term for the
modified gravity, and calculated the galaxy-galaxy power spectrum
semi-analytically.  (See also Ref. \cite{Sealfon:2004gz} for similar
argument.)  By comparing the predicted power spectrum with that of SDSS
galaxies \cite{Tegmark:2004uf}, we derived quantitative, although still
conditional, constraints on deviations from Newton's law of gravity.

In this paper, we improve our previous work by performing non-linear
cosmological simulations and by exploiting a higher-order statistic,
bispectrum. Since bispectrum is sensitive to clustering in the
non-linear regime, it is expected to provide complementary constraints
at mega-parsec scales to that obtained from power spectrum analysis.  We
use direct $N$-body simulations to test the accuracy of our
semi-analytic calculations and to reinforce our conclusions.

The rest of the paper is organized as follows.  Our model assumptions
are described in Sec. \ref{sec:Model}.  We derive power spectrum and
bispectrum from perturbation theory in modified Newtonian model in
Sec. \ref{sec:previous} and \ref{sec:2nd_perturbation}. We perform
$N$-body simulations and construct mock samples of volume-limited SDSS
galaxies for direct comparison with the observational data. Details of
the simulations are described in Sec. \ref{sec:Simulation}.  The results
of perturbation theory and the simulations are discussed in Sec.
\ref{sec:power} and \ref{sec:bi}.  Finally Sec. \ref{sec:summary}
concludes the present analysis.

\section{Model Assumptions}
\label{sec:Model}

In this section, we briefly summarize our model and a set of
assumptions.  Further details may be found in Paper I.

We consider a modified Newtonian model for which gravitational potential
is given by
\begin{equation}
\hspace*{-0.4cm}
\Phi (\br) = - G_{\rm N} \int dr'^3 \frac{\rho(\br')}{|\br-\br'|} 
\ldka 1 + \alpha \lka 1 - e^{- \frac{|\br - \br'|}{\lambda}} \rka
\rdka,
\label{eq:Model}
\end{equation}
where $G_{\rm N}$ denotes (conventional) Newton's constant of gravity.
The above model corresponds to Model II in Paper I, on which we focus
throughout the following analysis.  The deviation from the Newtonian
gravity in this model is characterized by two parameters, $\alpha$ and
$\lambda$; $\alpha$ is the dimensionless amplitude of the deviation and
$\lambda$ is the characteristic length scale.  Note that $\lambda$ is
defined in the proper length, rather than in the comoving length.

It is important to note that, although we consider deviations from
Newtonian gravity at mega-parsec scales, we still assume that the global
cosmic expansion is unaffected by the deviations. Namely, we assume that
general relativity is valid on horizon scales and thus the cosmic
expansion is described by the standard Friedmann equation.  Strictly
speaking, these two assumptions may be in conflict with modified gravity
models in general \cite{Deffayet:2001pu, Lue:2004rj, Alcaniz:2004kq,
Lue:2004za, Yamamoto:2006yv}.  To account for the existing data such as
SNeIa and CMB, however, the cosmic expansion law can hardly change in
practice.  This is why we adopt the conventional Friedmann equation even
in this analysis.  For the same reason, we use conventional matter
transfer function as initial condition of dark matter adopting the
background cosmology defined by the standard set of cosmological
parameters, $\Omega_{\rm m}$=0.3, $\Omega_{\rm b}$=0.04, $\Omega_\Lambda
= 0.7$, and the Hubble constant at present $h=0.7$ in units of 100km
$s^{-1}$ Mpc$^{-1}$.  See Paper I for further discussion on this point.

In order to make a direct comparison between the clustering of SDSS
galaxies and our model predictions, we need to assume a biasing relation
for the distribution of galaxies and that of matter.  For this purpose,
we adopt a commonly adopted deterministic relation:
\begin{equation}
 \delta_{\bk {\rm,galaxy}} = b_1 \delta_{\bk}
  + \frac{b_2}{2} \delta_{\bk}^2 , 
 \label{eq:bias}
\end{equation}
where $\delta_{\bk_{\rm, galaxy}}$ and $\delta_{\bk}$ are fractional
fluctuation of galaxy number and mass density, $b_1$ and $b_2$ are
linear and quadratic biasing parameters.  We consider only linear bias
(i.e., $b_2 = 0$) when we use power spectrum, whereas we consider both
$b_1$ and $b_2$ for analyses using bispectrum.  To derive constraints on
$\alpha$ and $\lambda$, $b_1$ is treated as a free parameter to adjust
the overall clustering amplitude.

\section{Power spectrum analysis}
\label{sec:previous}

In Fourier space, the modified gravitational potential in
Eq. \eqref{eq:Model} can be written as
\begin{equation}
\label{eq:pkModel}
\ldka \Delta_{\bx} \Phi ( \bx ) \rdka_{\bk}
	= 4 \pi G_{\rm N} a^2 \rb \ldka 1 + \alpha 
	 \frac{\lka \frac{a}{k \lambda} \rka^2}{1 + \lka \frac{a}{k
	 \lambda} \rka^2} \rdka \dbk,
\end{equation}
where $\bx$ is in the comoving coordinate, $\bk$ is the comoving
wave-number, and $a$ is the scale factor normalized unity at the present
epoch.

For the potential of Eq. \eqref{eq:pkModel}, the evolution equation for
density perturbations is written as
\begin{equation}
 \calDk \dbk^{(1)} = 0,
 \label{eq:3-37-1}
\end{equation}
with
\begin{gather}
\calDk \equiv \frac{d^2}{dt^2} + 2 H(a) \frac{d}{dt} - \calHk,
 \label{eq:3-35} \\
 \calHk \equiv \frac{3}{2} H^2(a) \Omega_m (a) \ldka
 1 + \alpha \frac{\lka \frac{a}{k \lambda} \rka^2}{1 + \lka \frac{a}{k
 \lambda} \rka^2} \rdka, \label{eq:3-8}
\end{gather}
where $H(a)$ is the Hubble parameter, and $\dbk^{(1)}$ denotes the
linear term in density fluctuations [see Eq. \eqref{eq:31-1} below].
Note that even the linear perturbation equation becomes dependent on $k$
in the modified gravity model.

Next, the linear power spectrum $P_{\rm L}(k)$ at present is given by
\begin{equation}
 P_{\rm L}(k; \alpha, \lambda) = A T^2(k) k^n
  \ldka \dbk^{(1)} (a=1; \alpha, \lambda) \rdka^2,
 \label{eq:linearP}
\end{equation}
where $T(k)$ is the matter transfer function, and $n$ is the spectral
index of the primordial power spectrum which we set to be unity.  We use
the fitting formula of Eisenstein and Hu \cite{Eisenstein:1997ik} for
$T(k)$.  It should be emphasized here that we fix the amplitude $A$ so
that the rms value of the top-hat mass fluctuations at 8$h^{-1}$Mpc,
$\sigma_8$, equals 0.9 \textbf{when $\alpha = 0$ and $n=1$}.  The actual
value of $\sigma_8$ in our modified gravity model may be slightly
different because of the factor $ \ldka \dbk^{(1)} (a=1; \alpha,
\lambda) \rdka^2$ in Eq. \eqref{eq:linearP}.  However, the difference in
the overall amplitude is unimportant because we have an additional
freedom to adjust the predicted amplitude via the biasing relation
[Eq. \eqref{eq:bias}].

In Paper I we used the Peacock-Dodds prescription \cite{Peacock:1996ci}
to convert the linear power spectrum to nonlinear one. It turned out
that in doing so we used incorrectly the growth factor $g(\Omega)$ and
the tilt of linear power spectrum $n_{\rm L}(k_{\rm L})$ given in the
case of Newtonian models. We made sure later that the above mistake did
not change the final power spectra very much as long as the
Peacock-Dodds prescription is valid.  In the present paper, we also
confirm the validity of the Peacock-Dodds approach in non-Newtonian
models using $N$-body simulations directly (see Sec. \ref{sec:power}).

\section{Perturbation theory and bispectrum}
\label{sec:2nd_perturbation}

In this section, we describe the second order perturbation theory and
its application to bispectrum.  The earlier formulation of cosmological
perturbation in the Newtonian model may be found in
\cite{Juszkiewicz:1981, Vishniac:1983, Suto:1990wf, Makino:1991rp}.
Bernardeau \cite{Bernardeau:2004ar} developed a formulation of second
order perturbation theory in non-Newtonian models.  We apply the method
to the modified potential in Eq. \eqref{eq:Model}.

The basic equations are given by
\begin{gather}
 \dot{\delta} + \frac{1}{a} \partial_i \ltka v^i \lka 1+ \delta \rka \rtka = 0,
 \label{eq:9} \\
 \dot{v}^i + \frac{1}{a} v^j \partial_j v^i + \frac{\dot{a}}{a} v^j = -
 \frac{1}{a} \partial_i \Phi, \label{eq:10}
\end{gather}
where the over-dot denotes the derivative with respect to time, $v^i(=a
\dot{x}^i)$ is the peculiar velocity, and $\Phi$ is the gravitational
potential.  We define velocity divergence:
\begin{equation}
 u(\bx, t) \equiv \partial_i v^i (\bx,t).
 \label{eq:12}
\end{equation}

Equations \eqref{eq:9} and \eqref{eq:10} in Fourier space reduce to
\begin{gather}
 a \dot{\delta}_\bk + \ubk = - \frac{1}{2(2 \pi)^3} \int d^3 p 
\ldka F(\bk, \bp) u_\bp \delta_\bq + F(\bk,\bq) u_\bq \delta_\bp \rdka
 \label{eq:27}\\
F(\bk, \bp) \equiv \frac{\bk \cdot \bp}{p^2}, \label{eq:29}
\\
\bq \equiv \bk - \bp,
\label{eq:20}
\end{gather}
and
\begin{gather}
 \dot{u}_\bk + Hu_\bk + a \calHk \dbk = - \frac{1}{(2 \pi)^3 a}
\int d^3 p G(\bk, \bp, \bq) u_\bp u_\bq,
 \label{eq:28} \\
G(\bk,\bp,\bq) \equiv \frac{k^2}{2 p^2} \frac{\bp \cdot \bq}{q^2}.
\label{eq:30}
\end{gather}

These equations can be solved recursively. Let us first decompose
$\delta_\bk$ and $u_\bk$ perturbatively,
\begin{gather}
 \dbk = \dbk^{(1)}+ \dbk^{(2)} + \dbk^{(3)} + \cdots,
 \label{eq:31-1}\\
 \ubk = \ubk^{(1)} + \ubk^{(2)} + \ubk^{(3)} + \cdots.
 \label{eq:31-2}
\end{gather}
Differentiating Eq. \eqref{eq:27} and substituting to Eq. \eqref{eq:28}
to eliminate $\ubk$, we obtain
\begin{equation}
 \calDk \dbk^{(n)} = \frac{\dot{A}^{(n)}}{a} + \frac{H}{a}A^{(n)} -
  \frac{B^{(n)}}{a},
 \label{eq:36}
\end{equation}
where $A^{(n)}$ and $B^{(n)}$ are the source terms of the $n$th-order:
\begin{gather}
 A^{(n)} \equiv - \frac{1}{2 (2\pi)^3} \int d^3 p \ldka
F(\bk, \bp) \sum_{i=1}^{n-1} u_{\bp}^{(i)} \delta_{\bq}^{(n-i)} + 
\lka \bp \leftrightarrow \bq \rka \rdka,
 \label{eq:32}\\
B^{(n)} \equiv - \frac{1}{(2 \pi)^3 a} \int d^3 p \ldka G(\bk, \bp, \bq)
\sum_{i=1}^{n-1} u_{\bp}^{(i)} u_{\bq}^{(n-i)} \rdka.
\label{eq:33}
\end{gather}

Consider first the lowest order, $n=1$.  Since $A^{(1)} = B^{(1)} = 0$,
Eq. \eqref{eq:36} reduces to
\begin{equation}
 \calDk \dbk^{(1)} = 0,
 \label{eq:37-1}
\end{equation}
which is equivalent to Eq. \eqref{eq:3-37-1}.  We denote the growing
mode of the solution of Eq. \eqref{eq:37-1} by $D_k^{(1)}(t)$.  Note
that, in non-Newtonian models, the solution $D_k^{(1)}(t)$ is generally
dependent on scale $k$, in contrast to the conventional Newtonian case.
The linear solution $\delta_\bk^{(1)}$ is given by
\begin{equation}
 \dbk^{(1)} = D_k^{(1)} \dini (\bk),
 \label{eq:37}
\end{equation}
where $\dini(\bk)$ is the initial fractional density.

The corresponding linear solution for $u_\bk$ is obtained from
Eq. \eqref{eq:27} as
\begin{equation}
 \ubk^{(1)} = - a \dot{\delta}_\bk^{(1)} = - a \dot{D}_k^{(1)} \dini
 (\bk).
\label{eq:38}
\end{equation}

Solutions at the next order, $n=2$, are more complicated.
Eq. \eqref{eq:36} for $n=2$ is written explicitly as
\begin{widetext}
\begin{gather}
 \calDk \dbk^{(2)} = \frac{1}{(2 \pi)^3} \int d^3 p d^3 q \delta^D (\bp
 + \bq - \bk) \ltka \dini (\bp) \dini (\bq)
 \ldka S_0(p,q,t) \calP_0(\mu) +
 S_1(p,q,t) \calP_1(\mu) + S_2(p,q,t) \calP_2(\mu) \rdka \rtka,
 \label{eq:49}\\
S_0(p,q,t) = \lka \frac{\calH_p}{2} + \frac{\calH_q}{2} \rka
D_p^{(1)} D_q^{(1)} + \frac{4}{3} \dot{D}_p^{(1)} \dot{D}_q^{(1)},
\label{eq:50-1}\\
S_1(p,q,t) = \lka \frac{\calH_p}{2} \frac{q}{p} +
 \frac{\calH_q}{2} \frac{p}{q} \rka D^{(1)}_p D^{(1)}_q
+ \lka \frac{q}{p} + \frac{p}{q} \rka \dot{D}_p^{(1)} \dot{D}_q^{(1)},
\label{eq:50-2}\\
S_2(p,q,t) = \frac{2}{3} \dot{D}_p^{(1)} \dot{D}_q^{(1)},
\end{gather}
\end{widetext}
where $\delta^D(\bk)$ is the Delta function and $\mathcal{P}_l (\mu)$
are the Legendre polynomials:
\begin{gather} 
\calP_0(\mu) = 1, \quad \calP_1(\mu)= \mu, \quad \calP_2(\mu)=
 \frac{1}{2} \lka 3 \mu^2 - 1 \rka,
 \label{eq:47}\\
\mu \equiv \frac{\bp \cdot \bq}{pq}.
\label{eq:47-1}
\end{gather}
Equation \eqref{eq:49} has an implicit solution of the form:
\begin{gather}
 \dbk^{(2)} = \frac{1}{(2 \pi)^3} \int d^3 p d^3 q \delta^D (\bp
 + \bq - \bk) \ldka \dini (\bp) \dini (\bq) \times \right. \notag \\
\left. \ltka T_0(p,q,t) \calP_0(\mu) +
 T_1(p,q,t) \calP_1(\mu) + T_2(p,q,t) \calP_2(\mu) \rtka \rdka,
 \label{eq:49_1}
\end{gather}
where the functions $T_i (p,q,t)$ satisfy
\begin{equation}
 \calD_{|\bp + \bq|} T_i (p,q,t) = S_i (p,q,t)
\quad \text{for $i=0,1, 2$}.
 \label{eq:56}
\end{equation}
%%%%%%%%%%%%%%%%%%%%%%%%
We note that expressions for the second-order solutions given in
\cite{Bernardeau:2004ar} contain some typographical errors which are
corrected in our above expressions.
%%%%%%%%%%%%%%%%%%%%%%%%

These results enable us to compute the bispectrum in the leading order.
The bispectrum is defined as
\begin{equation}
 \langle \delta (\bk_1) \delta (\bk_2) \delta (\bk_3) \rangle
= \lka 2 \pi \rka^3 B(\bk_1, \bk_2, \bk_3) 
\delta^D (\bk_1 + \bk_2 + \bk_3).
 \label{eq:defB}
\end{equation}
The leading-order terms of the left-hand-side of the above equation are
given by
\begin{gather}
 \langle \delta (\bk_1) \delta (\bk_2) \delta (\bk_3) \rangle
= \langle \delta^{(2)} (\bk_1) \delta^{(1)} (\bk_2) \delta^{(1)} (\bk_3) 
\rangle \notag \\
+ \textit{cyc.}(1,2,3).
 \label{eq:treeB}
\end{gather}
Therefore the bispectrum reduces to
\begin{gather}
 B(\bk_1, \bk_2, \bk_3) = 
2 D^{(1)}_{k_1} D^{(1)}_{k_2} \ldka \sum_{i=0}^{2}
  T_i(k_1,k_2,t) \calP_i(\bk_1, \bk_2) \rdka \notag \\
  \times P_{\rm ini}(k_1) P_{\rm ini}(k_2)
+ cyc.(1,2,3) ,
 \label{eq:59}
\end{gather}
where $P_{\rm ini} (k) \equiv \langle | \dini(\bk) |^2 \rangle $.  In
what follows, we write the bispectrum simply as $B(\bk_1, \bk_2)$
adopting the condition of $\bk_3 = -\bk_2 - \bk_1$
[Eq. \eqref{eq:defB}].

To compute the bispectrum, we solve Eq. \eqref{eq:56} numerically for
each pair of $(\bp, \bq)$, together with the linear perturbation
equation \eqref{eq:37-1}.  At sufficiently early epochs ($z_i \gg 1$),
$D^{(1)}_k (z_i)$ is simply given by the growth rate in the Newtonian
case (see Paper I).  Similarly, $T_i$ are given by
\begin{gather}
 T_{0}(p,q,z_i) = \frac{17}{21} (1+z_i)^{-2}, \label{eq:EdS_F0} \\
 T_{1}(p,q,z_i) = \frac{1}{2} \lka \frac{q}{p} + \frac{p}{q} \rka (1+z_i)^{-2}
, \label{eq:EdS_F1} \\
 T_{2}(p,q,z_i) = \frac{4}{21} (1+z_i)^{-2}. \label{eq:EdS_F2}
\end{gather}

\section{Simulation and observational data}
\label{sec:Simulation}

\subsection{N-body Simulations}\label{}%%-----------------------------

We use the cosmological $N$-body solver TPM-1.1 \cite{Bode:2003ct} in
its PM-only mode.  We run six realizations each for simulation box-sizes
of $L_{\rm box} = 500 h^{-1}$Mpc, and $1000h^{-1}$Mpc with the following
parameters: $\alpha = $ $-1.0, -0.8, -0.5, -0.2, 0.0, 0.2, 0.5, 0.8,$
and $1.0$, $\lambda = 2, 5, 8, 10, 12, 15, 20,$ and $30 h^{-1}$Mpc.  We
use the fitting formula for the matter transfer function, equation (28)
$\sim$ (31) of the ref. \cite{Eisenstein:1997ik}, that ignores the
baryon acoustic oscillation effect.  We start the simulations at
$z=50$. All the simulations employ $N_p = 128^3$ particles.

To simulate structure formation in the non-Newtonian model, we need to
modify the Green function of the Laplacian, $\hat{\mathcal{G}}$.  For a
density field $\hat{\rho}$ defined on a three-dimensional wave-number
grid $(p,q,r)$, the gravitational potential in real space is evaluated
to be
\begin{equation}
 \phi (l, m, n) = \sum_{p,q,r = 0}^{M-1}
\hat{\mathcal{G}}_{p,q,r} \hat{\rho}_{p,q,r}
\exp \left[2 \pi i (pl + qm + rn)/M \right],
 \label{eq:Efs-1}
\end{equation}
where $l,m,n$ are position integers in real space with $M$ being the
number of grids per dimension (we follow the notation in Efstathiou
{\it et al.} \cite{Efstathiou:1985re}).

The Green function in the original TPM code that assumes the
conventional Newtonian gravity is given by
\begin{gather}
\hat{\mathcal{G}}_{p,q,r}^{\rm old} = 
\begin{cases}
0, \qquad \qquad l=m=n=0; \\
- \pi/ \left\{ M^2 \left[ \sin^2 (\pi p / M)
\right. \right.  \\
\left. \left.
\quad + \sin^2 (\pi q / M ) + \sin^2 (\pi r / M) \right]\right\} 
\quad \text{otherwise;}
\end{cases}
\label{eq:Efs-2}
\end{gather}
which is derived from the seven-point finite-difference approximation.

Taking account of the scale-dependence in Eq. \eqref{eq:pkModel}, we
correct the Green function for the modified Newtonian model:
\begin{equation}
\hat{\mathcal{G}}_{p,q,r}^{\rm new} = 
\hat{\mathcal{G}}_{p,q,r}^{\rm old} \times
\left[ 1 + \alpha \frac{ (\frac{a}{k \lambda})^2}{1 + (\frac{a}{k
 \lambda})^2} \right].
\label{eq:Green}
\end{equation}
Note that $k$ in Eq. \eqref{eq:Green} needs to be given in the form,
consistently with the Green function itself, as
\begin{gather}
 k(p,q,r) = 
\frac{M}{\pi} \left\{ \left[ \sin^2 (\pi p / M)
+ \sin^2 (\pi q / M )  \right. \right.
\notag \\
\left. \left. \qquad \quad + \sin^2 (\pi r / M) \right]\right\}^{1/2}
 \times \frac{2 \pi}{L_{\rm box}}.
 \label{eq:wavenumber}
\end{gather}
We use the above Green function, evolve the system from $z=50$ to $0$,
and make mock galaxy samples in the manner described in the next
subsection.

%%%%%%%%%%%%%%%%%%%%%%%%%%%%%%%%%%%%%%%%%%%%%%%%%%%%%%%%%%%%%%%%%%%%%%%
% Fig. 1
%%%%%%%%%%%%%%%%%%%%%%%%%%%%%%%%%%%%%%%%%%%%%%%%%%%%%%%%%%%%%%%%%%%%%%%
\begin{figure*}
\begin{tabular}{cc}
      \resizebox{80mm}{!}{\includegraphics{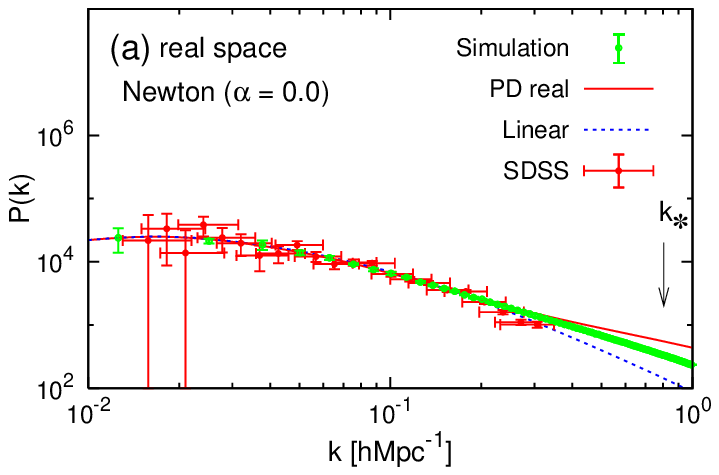}}
 &
      \resizebox{80mm}{!} {\includegraphics{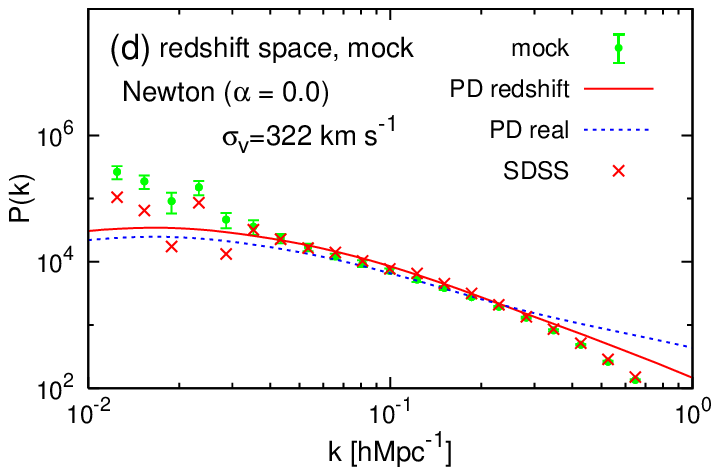}}
 \\
      \resizebox{80mm}{!}{\includegraphics{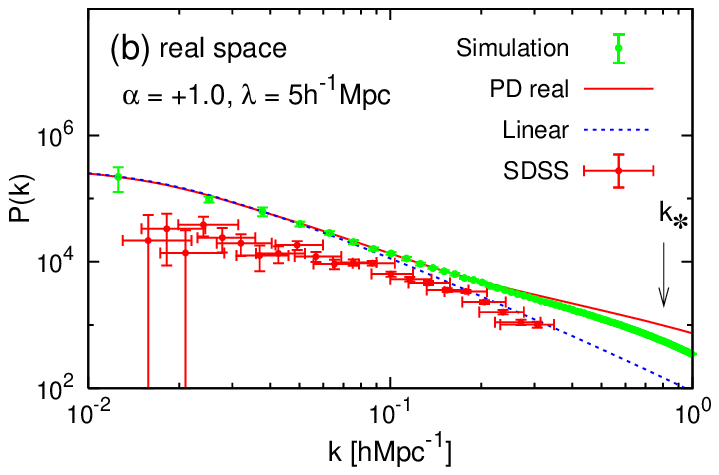}}
 &
      \resizebox{80mm}{!} {\includegraphics{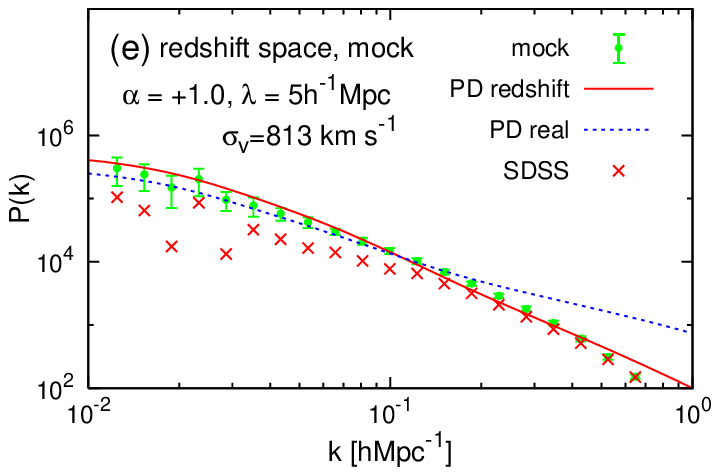}}
 \\
      \resizebox{80mm}{!}{\includegraphics{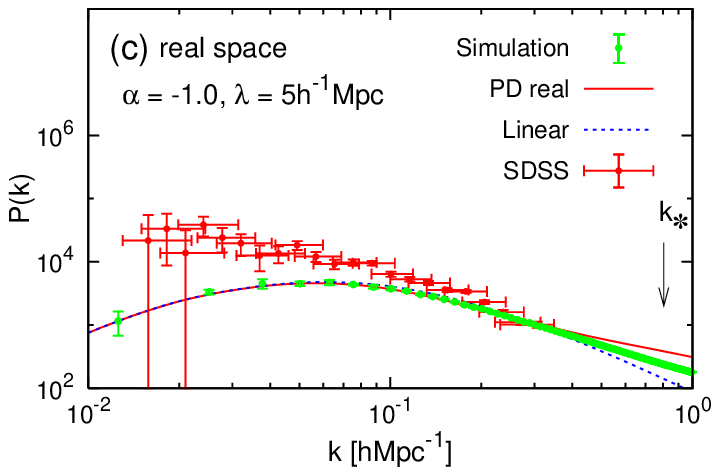}}
 &
      \resizebox{80mm}{!} {\includegraphics{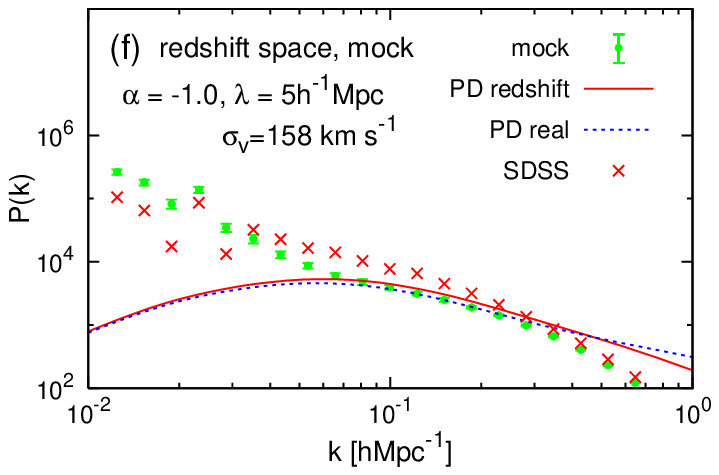}}
    \end{tabular}
\caption{\label{fig:PD} The panels of the left side show the power
 spectra in real space.  The adopted model parameters are (a) Newtonian
 ($\alpha = 0.0$), (b)$\alpha = +1.0$ and $\lambda = 5h^{-1}$Mpc, (c)
 $\alpha = -1.0$ and $\lambda = 5h^{-1}$Mpc.  Dots with vertical and
 horizontal error-bars are the power spectrum of SDSS galaxies from
 Tegmark {\it et al}. \cite{Tegmark:2004uf}.  Dots with only vertical
 error-bars indicate results of N-body simulations.  Dotted and solid
 lines are linear and non-linear power spectrum, respectively.  We
 denote by $k_*$ the length scale of the mean inter-particle separation
 in our simulations, which is given by $k_{*} = 0.5 \cdot 2 \pi \cdot
 N_{\rm p}^{1/3}/ L_{\rm box}$. The simulation results are reliable at $k <
 k_*$.  In the right panels, we plot the power spectra for mock
 ``galaxies'' generated from our simulation.  The parameters for (d),
 (e), (f) are the same as for (a), (b), (c), respectively.  The dotted
 and solid lines in the right panels are non-linear power spectrum in
 real space (, which are the same as solid lines in the left panels) and
 redshift space.  $\sigma_v$ means the one-dimensional velocity
 dispersion calculated from simulation data.  Cross symbols are the
 power spectra of the volume-limited sample of SDSS galaxies}
\end{figure*}
%%%%%%%%%%%%%%%%%%%%%%%%%%%%%%%%%%%%%%%%%%%%%%%%%%%%%%%%%%%%%%%%%%%%%%%

\subsection{Observational data and mock samples}
\label{subsec:mock}%%-----------------------------

For definiteness, we choose a volume-limited sample of SDSS galaxies
whose $r$-band magnitude is in the range of (-21.0, -20.0) from those
described in Hikage {\it et al}. \cite{Hikage:2005ia}.  The redshift
range is $0.044 < z < 0.103$, the survey volume, $V_{\rm samp}$, is
$9.20 \times 10^6 (h^{-1}{\rm Mpc})^3$, and the total number of galaxies
is 44,636.  We made sure that using the other volume-limited samples
with different magnitude ranges \cite{Hikage:2005ia} does not
significantly affect the results of our analysis below.

We generate 24 mock catalogues from our $N$-body simulation data.
The mock catalogues take into account various observational effects such
as survey geometry, the number density, and redshift distortion
(peculiar velocities of simulations particles are assigned to the mock
galaxies ) \cite{Hikage:2005ia}.  In order to account for the effect of
survey geometry, we distribute random particles within the survey volume
and correct for the boundary effect following the prescription of
Feldman, Kaiser and Peacock \cite{FKP}.  We subtract fluctuations of the
random particles which are within the survey volume, $\delta_{\bk, {\rm
random}}$:
\begin{equation}
\label{eq:delta_mock}
\tilde{\delta}_{\bk} = \delta_{\bk,{\rm data}} - \delta_{\bk,{\rm random}}.
\end{equation}
While this prescription is fairly empirical and may not completely
account for the effect of the survey geometry, it yields a robust
estimate at scales of our main interest here, $k \sim 0.1 h{\rm
Mpc}^{-1}$.  When we calculate the power spectrum and bispectrum for
SDSS galaxies and the mock catalogues, we use the above ``corrected''
density, $\tilde{\delta}_{\bk}$.

\section{Constraints from power spectrum}
\label{sec:power}

We first compare the power spectra used the Peacock-Dodds prescription
and those from numerical simulations.  In Fig. \ref{fig:PD}, we plot the
mass power spectra in real space (left panels) and in redshift space
(right panels).  The predictions from perturbation theory agree well
with the results of $N$-body simulations..  Note that in the Newtonian
case, the predicted power spectra with $b_1=1$ are already in reasonable
agreement with the observed power spectrum of SDSS galaxies.  Our
simulation results are also consistent with those of Stabenau and Jain
\cite{Stabenau:2006td}.

The panels on the right side in Fig. \ref{fig:PD} show the power spectra
of our mock ``galaxies''.  In each panel, the dotted line indicates the
non-linear power spectrum in real space, which is the same in the
corresponding left panel and shown for comparison. The redshift-space
power spectrum of the SDSS volume-limited sample is shown by cross
symbols.  To include effects of redshift space distortion in our model,
we use the formula derived of Magira, Jing and Suto \cite{Magira:1999bn}
[equation (12) in their paper].  On linear scales ($k < 0.1
h$Mpc$^{-1}$), the Kaiser effect is clearly seen as an enhanced power
with respect to the real space power spectrum. It is worth mentioning
that the plotted power spectra show substantial variations on the
largest scales ($k < 0.03 h {\rm Mpc}^{-1}$), which are presumably due
to the somewhat complex survey geometry.

%%%%%%%%%%%%%%%%%%%%%%%%%%%%%%%%%%%%%%%%%%%%%%%%%%%%%%%%%%%%%%%%%%%%%%%
% Fig. 2
%%%%%%%%%%%%%%%%%%%%%%%%%%%%%%%%%%%%%%%%%%%%%%%%%%%%%%%%%%%%%%%%%%%%%%%
\begin{figure}[htbp]
\begin{tabular}{c}
      \resizebox{80mm}{!}
 {\includegraphics{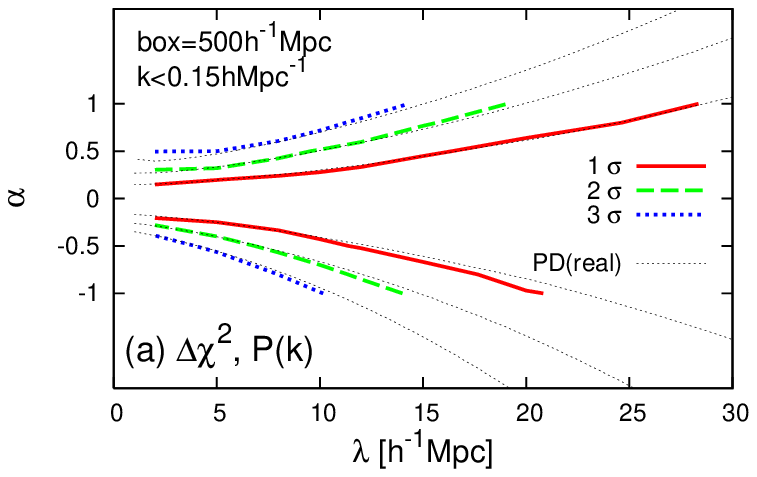}}
\\
      \resizebox{80mm}{!}
 {\includegraphics{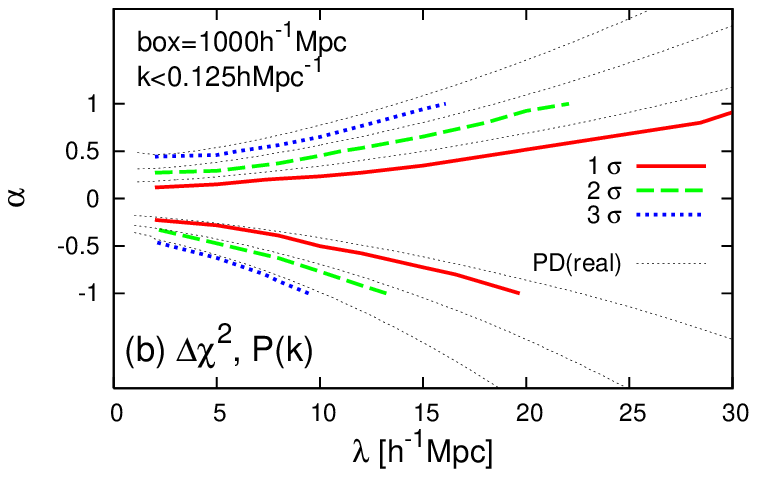}}
\\
      \resizebox{80mm}{!}
 {\includegraphics{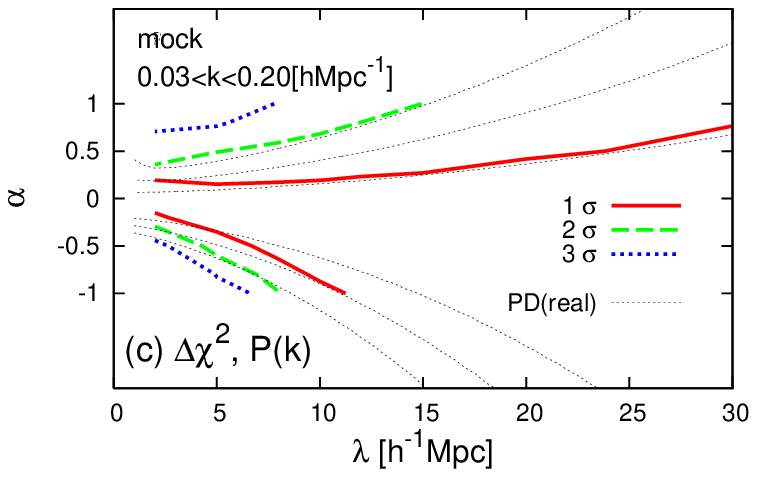}}
\end{tabular}
\caption{ \label{fig:simP_constraint} Constraints on $\alpha$ and
$\lambda$ obtained using (a) simulations with $L_{\rm box}=500
h^{-1}$Mpc, (b) simulations with $L_{\rm box}=1000 h^{-1}$Mpc, and (c)
mock galaxy catalogues.  The range of $k$ used to compute $\chi^2$ is
indicated in each panel.  Thick solid, dotted, thick dotted lines
indicate the limits at 1$\sigma$, 2$\sigma$ and 3$\sigma$ confidence
levels. For comparison, we plot the corresponding 1$\sigma$, 2$\sigma$
and 3$\sigma$ confidence levels using the real-space Peacock-Dodds
prediction in thin dotted lines. }
\end{figure}
%%%%%%%%%%%%%%%%%%%%%%%%%%%%%%%%%%%%%%%%%%%%%%%%%%%%%%%%%%%%%%%%%%%%%%%

To derive constraints on $\alpha, \lambda$ using the calculated power
spectra, we apply the $\Delta \chi^2$ statistic.  We treat the linear
bias parameter $b_1$ as a free parameter in order to adjust the overall
amplitudes of the power spectra between the predictions/simulations and
the SDSS data. This normalization allows us to use the {\it shape} of
the power spectra to detect possible deviations from the Newtonian case.

We calculate $\chi^2$ as 
\begin{equation}
 \chi^2 \equiv \sum_i \frac{[P(k_i) - P_{\rm SDSS}(k_i)]^2}{\sigma^2 (k_i)},
\label{eq:chi2}
\end{equation}
where $P_{\rm SDSS}(k_i)$ is the SDSS galaxy power spectrum.  We use the
predicted power spectra $P(k_i)$ and the variance of the SDSS data,
$\sigma^2(k_i)$, to calculate $\chi^2$ in real space, while for the same
analysis in redshift space, we use those power spectra with the variance
of mock galaxy samples to represent the cosmic variance in redshift
space.

We compute the relative confidence level of $\alpha$ and $\lambda$ with
respect to their best-fit values assuming that
\begin{align}
\label{eq:chi}
\Delta \chi^2 (\alpha,\lambda) \equiv
	\chi^2(\alpha,\lambda,b_{*, \text{local~min}}) 
  - \chi^2(\alpha_{\rm min},\lambda_{\rm min},b_{*, \text{min}}) 
\end{align}
follows the $\chi^2$distribution for 2 degrees of freedom.  In
Eq. (\ref{eq:chi}), $\alpha_{\rm min}$, $\lambda_{\rm min}$ and $b_{*,
\text{min}}$ denote their best-fit values which globally minimize the
value of $\chi^2$, while $b_{*,\text{local~min}}$ is the value that
minimizes the $\chi^2$ for a given set of values of $\alpha$ and
$\lambda$.

Figure \ref{fig:simP_constraint} shows the contours of $\Delta \chi^2
(\alpha, \lambda)$. The results from $N$-body simulations in real space
are shown in panel (a) and (b). These differ only in the simulation box
size, 500 $h^{-1}$Mpc for (a) and 1000 $h^{-1}$Mpc for (b). Hence the
range of $k$ used to derive constraints is slightly different, We also
show the result from the real-space Peacock-Dodds prediction by thin
dotted lines using the the same range of $k$ consistently with the
simulations.  Clearly, the results of the perturbation theory and that
of our numerical simulations are consistent with each other, putting
quite similar constraints on $\alpha$ and $\lambda$.

The bottom panel (c) in Figure \ref{fig:simP_constraint} shows the
constraints from our mock galaxy samples in redshift space.  The range
of $k$ used in the analysis is $0.03 < k < 0.20 h$Mpc$^{-1}$.  The
constraint is slightly less tight than those from perturbation theory
and $N$-body simulations. This is mainly because we discard the data
points at large scales $k \sim 0.01h$Mpc$^{-1}$ where the deviations from the
Newtonian case are most significant.  Nevertheless models with $|\alpha|
> 1$ are still excluded at a 2-3$\sigma$ confidence level for $\lambda
\sim 10 h^{-1}{\rm Mpc}$. For reference, we also plot the contours based
on the real-space Peacock-Dodds prediction by thin dotted lines.

\section{Constraints from bispectrum}
\label{sec:bi}

We further derive constraints on the modified Newtonian model extending
the analysis to the three-point statistics. Specifically we use
(conventional) bispectrum, $B(\bk_1, \bk_2)$, defined in Eq.
\eqref{eq:defB}, and reduced bispectra $Q$ and $p^{(3)}$ defined as
\begin{equation}
 Q(\bk_1, \bk_2) = 
  \frac{B(\bk_1, \bk_2)}{P(k_1) P(k_2) + P(k_2) P(k_3) + P(k_3) P(k_1)},
 \label{eq:defQ}
\end{equation}
and 
\begin{equation}
 p^{(3)}(\bk_1, \bk_2) = 
  \frac{B(\bk_1, \bk_2)}{\sqrt{V_{\rm samp} P(k_1) P(k_2) P(k_3)}},
 \label{eq:defp3}
\end{equation}
where $\bk_3 \equiv - \bk_1 - \bk_2$, $k_i = |\bk_i|$, and $V_{\rm
samp}$ is the sampling volume. The latter quantity $p^{(3)}$ is the
probability density function of phase sum for a density field,
$\theta_{\bk_1} + \theta_{\bk_2} + \theta_{\bk_3}$ [$\delta_{\bk} =
|\delta_{\bk}| \exp(i \theta_\bk)$], studied in Matsubara
\cite{Matsubara:2003te} and Hikage {\it et al}.
\cite{Hikage:2003kr,Hikage:2005ia}.  In this paper, we consider only
isosceles triangles in $k$-space that satisfy the relation $k \equiv k_1 = k_2$
with angle $\varphi$ defined as
\begin{equation}
 \varphi = \cos^{-1} \lka \frac{\bk_1 \cdot \bk_2}{k_1 k_2} \rka.
 \label{eq:defphi}
\end{equation}

In the following analysis, we use $p^{(3)}$ to give constraints on the deviation from Newtonian gravity.
This is because $p^{(3)}$ consists only of Fourier-phase informations and thus
their constraints have good complementarity with those from $P(k)$, which is defined as
the square of the Fourier amplitudes.

%%%%%%%%%%%%%%%%%%%%%%%%%%%%%%%%%%%%%%%%%%%%%%%%%%%%%%%%%%%%%%%%%%%%%%%
% Fig. 3
%%%%%%%%%%%%%%%%%%%%%%%%%%%%%%%%%%%%%%%%%%%%%%%%%%%%%%%%%%%%%%%%%%%%%%%
\begin{figure*}
\begin{tabular}{cc}
      \resizebox{80mm}{!}{\includegraphics{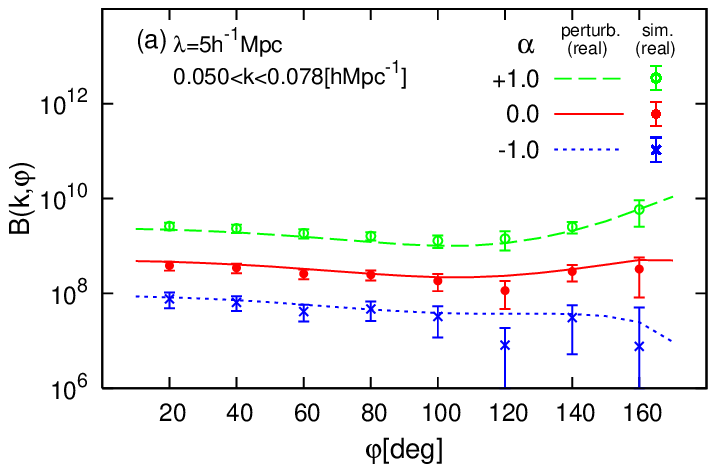}}
 &
      \resizebox{80mm}{!}{\includegraphics{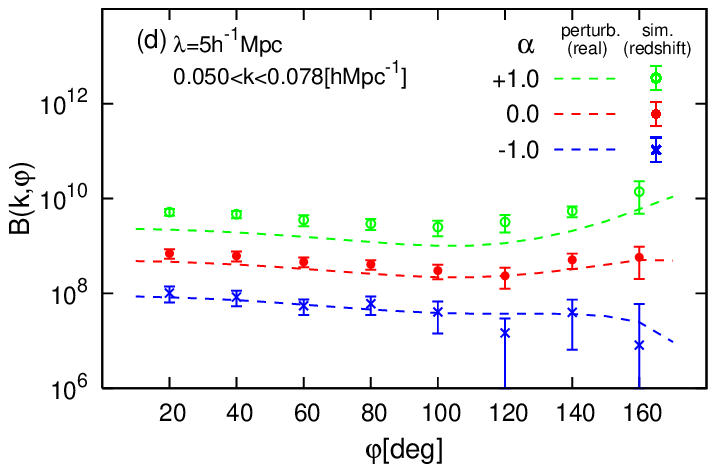}}
 \\
      \resizebox{80mm}{!}{\includegraphics{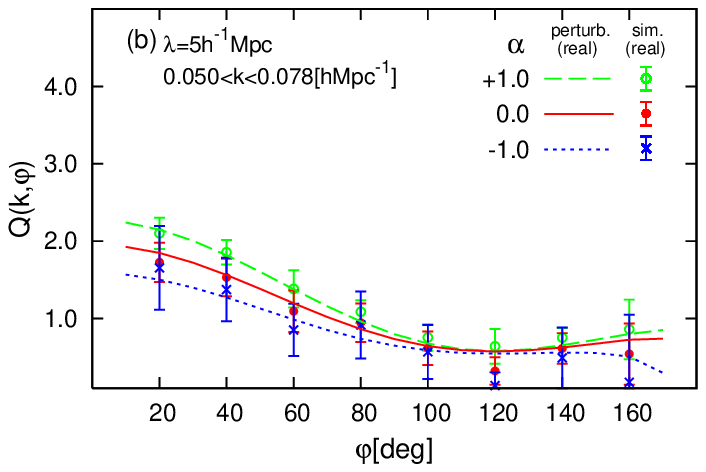}}
 &
      \resizebox{80mm}{!}{\includegraphics{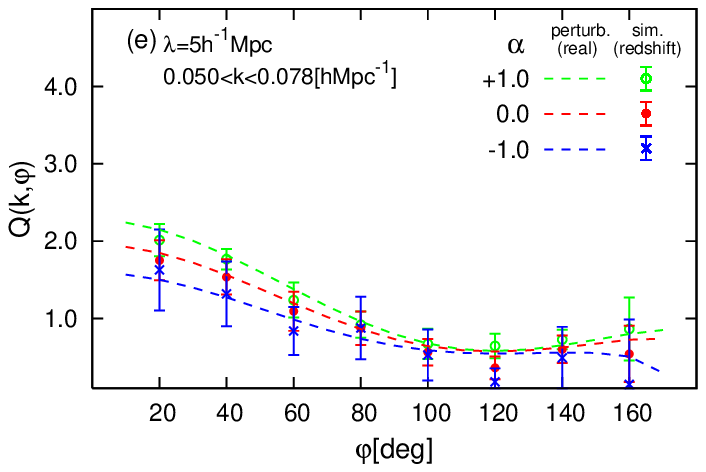}}
 \\
      \resizebox{80mm}{!}{\includegraphics{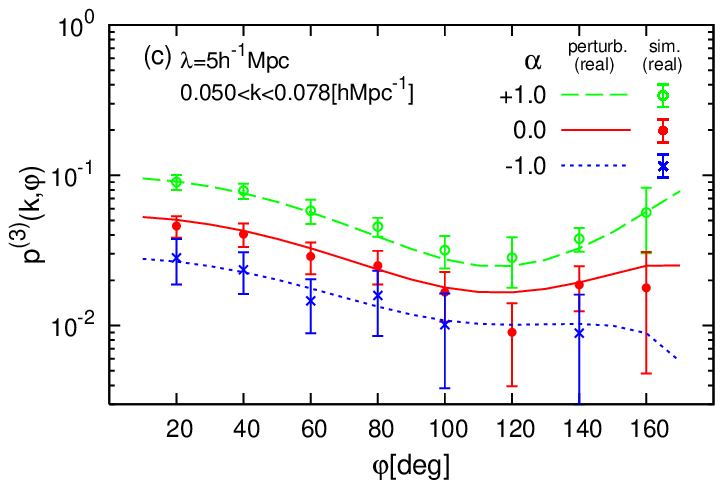}}
 &
      \resizebox{80mm}{!}{\includegraphics{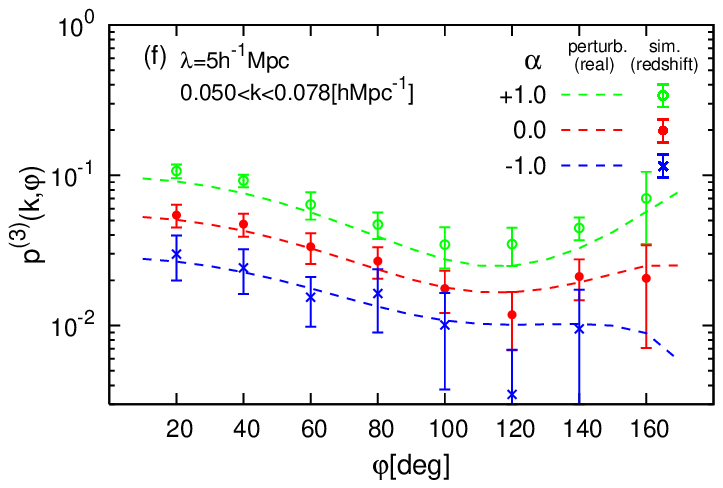}}
    \end{tabular}
\caption{\label{fig:bi_full_a} Bispectra $B(k)$, $Q(k)$, and
 $p^{(3)}(k)$ from top to bottom as a function of $\varphi$ measured in
 real space (left panels) and in redshift space (right panels).  The
 range of $k \equiv |\bk_1| = |\bk_2|$ is indicated in each panel. The
 value of $\lambda$ is fixed as $5h^{-1}$Mpc. The dashed, solid, dotted
 lines show the perturbation predictions in real space for $\alpha =
 +1.0$, $0.0$ (Newtonian) and $-1.0$, respectively.  Symbols with
 error-bars show the results of simulations.  Open circle, filled
 circle, cross symbol also mean $\alpha = +1.0$, $0.0$(Newton) and
 $-1.0$, respectively.  }
\end{figure*}
%%%%%%%%%%%%%%%%%%%%%%%%%%%%%%%%%%%%%%%%%%%%%%%%%%%%%%%%%%%%%%%%%%%%%%%

%%%%%%%%%%%%%%%%%%%%%%%%%%%%%%%%%%%%%%%%%%%%%%%%%%%%%%%%%%%%%%%%%%%%%%%
% Fig. 4
%%%%%%%%%%%%%%%%%%%%%%%%%%%%%%%%%%%%%%%%%%%%%%%%%%%%%%%%%%%%%%%%%%%%%%%
\begin{figure*}
\begin{tabular}{cc}
      \resizebox{80mm}{!}{\includegraphics{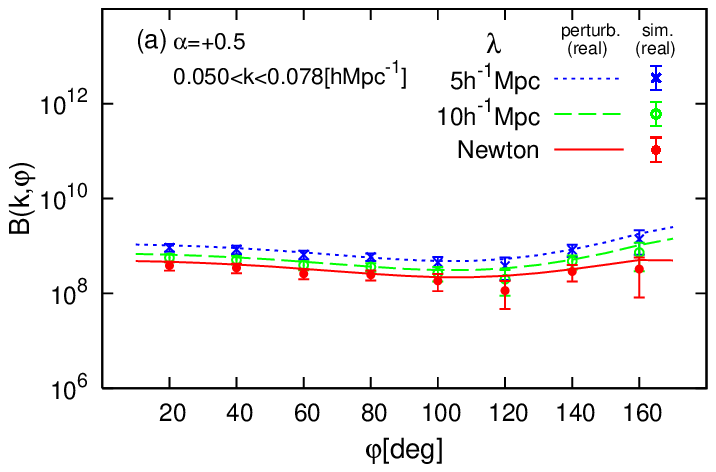}}
 &
      \resizebox{80mm}{!}{\includegraphics{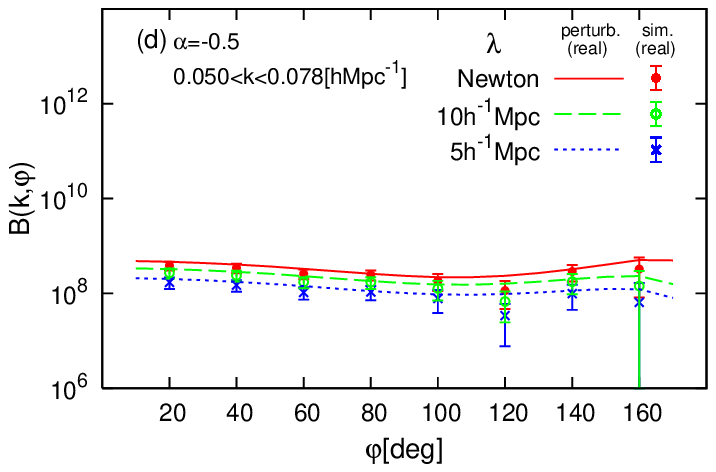}}
 \\
      \resizebox{80mm}{!}{\includegraphics{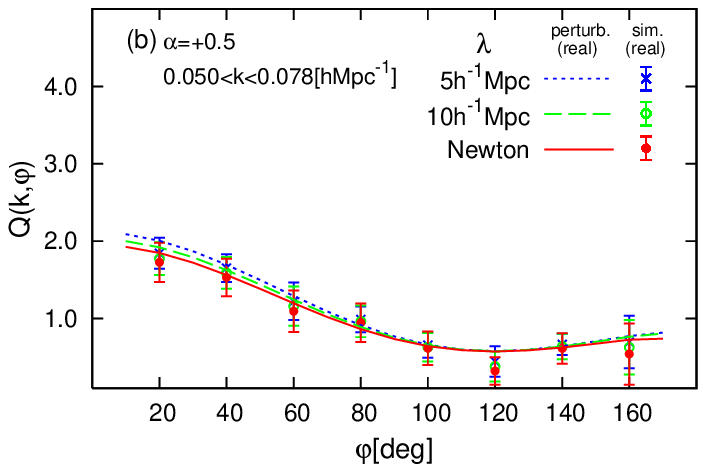}}
 &
      \resizebox{80mm}{!}{\includegraphics{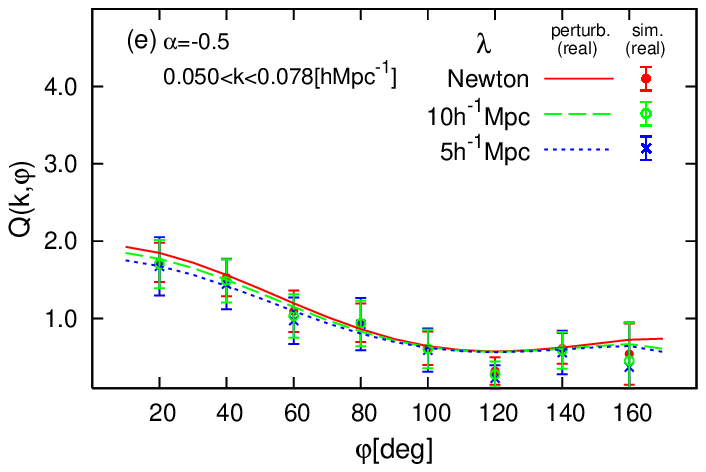}}
 \\
      \resizebox{80mm}{!}{\includegraphics{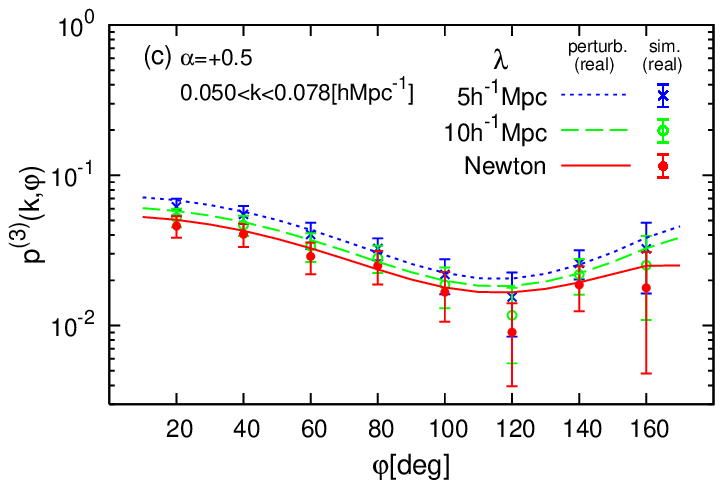}}
 &
      \resizebox{80mm}{!}{\includegraphics{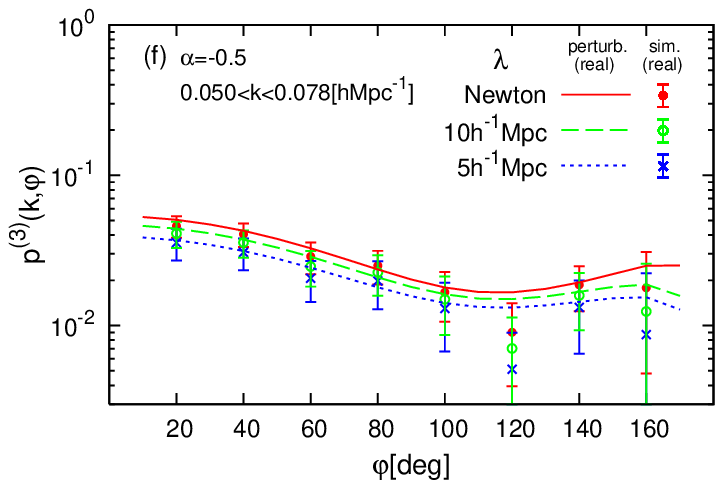}}
    \end{tabular}
\caption{\label{fig:bi_full_l} Bispectra $B(k)$, $Q(k)$, and
$p^{(3)}(k)$ from top to bottom as a function of $\varphi$ measured in
real space; {\it right:} $\alpha=0.5$, {\it left:} $\alpha=-0.5$.  The
dotted, dashed, and solid lines show the perturbation predictions in real
space for $\lambda = 5h^{-1}$Mpc, $\lambda = 10h^{-1}$Mpc, and
$\lambda=\infty$(Newtonian), respectively, while symbols indicate the
 corresponding simulation results.  }
\end{figure*}
%%%%%%%%%%%%%%%%%%%%%%%%%%%%%%%%%%%%%%%%%%%%%%%%%%%%%%%%%%%%%%%%%%%%%%%

%%%%%%%%%%%%%%%%%%%%%%%%%%%%%%%%%%%%%%%%%%%%%%%%%%%%%%%%%%%%%%%%%%%%%%%
% Fig. 5
%%%%%%%%%%%%%%%%%%%%%%%%%%%%%%%%%%%%%%%%%%%%%%%%%%%%%%%%%%%%%%%%%%%%%%%
\begin{figure}[htbp]
\begin{tabular}{c}
      \resizebox{80mm}{!}
 {\includegraphics{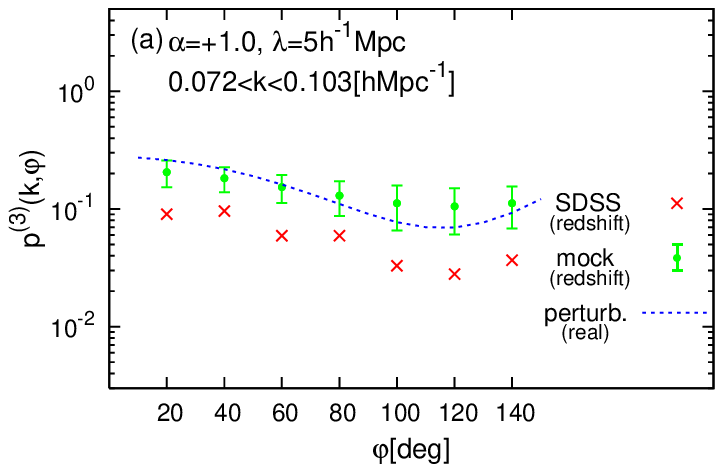}}
\\
      \resizebox{80mm}{!}
 {\includegraphics{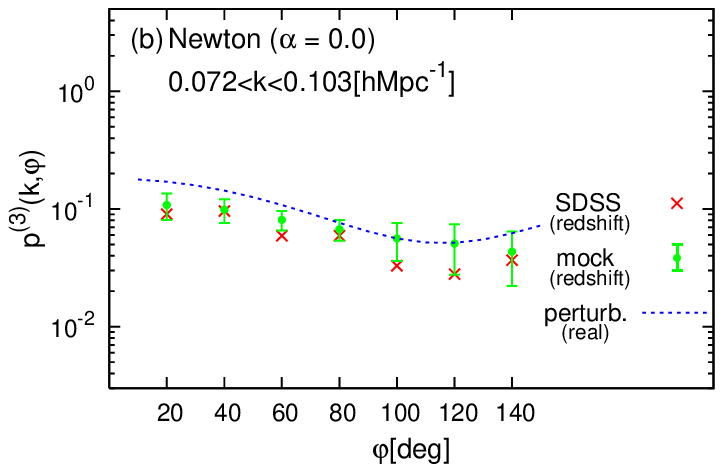}}
\\
      \resizebox{80mm}{!}
 {\includegraphics{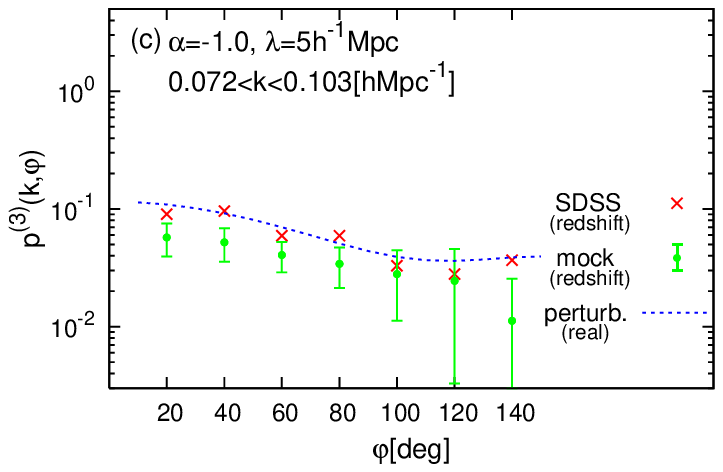}}
\end{tabular}
\caption{ \label{fig:mock_p3_-21.0} Bispectra $p^{(3)}$ for SDSS
galaxies (crosses), mock galaxy samples (solid circles with error bars),
and perturbation theory predictions in real space (dotted line):
(a)$\alpha = +1.0$ and $\lambda = 5 h^{-1}$Mpc, (b) $\alpha = 0.0$
(Newtonian), (c) $\alpha = -1.0$ and $\lambda = 5 h^{-1}$Mpc.}
\end{figure}
%%%%%%%%%%%%%%%%%%%%%%%%%%%%%%%%%%%%%%%%%%%%%%%%%%%%%%%%%%%%%%%%%%%%%%%

\subsection{Linear bias model with $b_2=0$}

Let us consider first linear bias model [$b_2 = 0$ in
Eq. \eqref{eq:bias}].  Figure \ref{fig:bi_full_a} plots the bispectra $B,
Q, p^{(3)}$ in real space (left panels) and in redshift space (right
panels) for $L_{\rm box}=$500 $h^{-1}$Mpc simulations.  The survey
volume is set to be $(L_{\rm box})^3$ in Eq. \eqref{eq:defp3}.  The
bispectra at small $\varphi$ are dominated by various nonlinear effects,
whereas there are substantial uncertainties at large $\varphi$ because
of the small number of Fourier modes sampled. Given those, the agreement
between predictions from perturbation theory (dashed lines) and $N$-body
simulation data (solid circles with error-bar) is very satisfactory.

The right panels of Fig. \ref{fig:bi_full_a} shows the bispectra in
redshift space. There, the results from our mock samples are shown by
symbols with error bars.  For comparison, we also show the results from
perturbation theory in {\it real space}.  In Fig. \ref{fig:bi_full_a}(d),
Kaiser effect is clearly seen as a enhance at small $\varphi$.

We further examine the dependence of the bispectra on $\lambda$.
Figure \ref{fig:bi_full_l} compares the bispectra for different values of
$\lambda$. We have set $\alpha$ = 0.5 (left panels) and $\alpha = -0.5$
(right panels).

Figure \ref{fig:mock_p3_-21.0} shows $p^{(3)}(k)$ for the volume-limited
SDSS catalogue and for our mock samples at $k$ in the range of $0.072h
{\rm Mpc}^{-1} <k < 0.103 h {\rm Mpc}^{-1}$.  They have a very
similar shape, but their amplitude depends systematically on the value
of $\alpha$, the degree of deviations from the Newtonian case.

\begin{figure}[htbp]
\begin{tabular}{c}
      \resizebox{80mm}{!}
 {\includegraphics{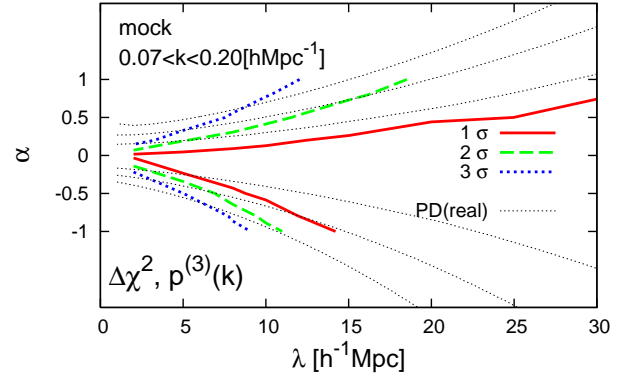}}
\end{tabular}
\caption{ \label{fig:mock_constraint_k0.07-0.20} Constraints on the
 $\alpha-\lambda$ plane from the $p^{(3)}$ analysis assuming $b_2=0$.
 The range of $k$ is from 0.07 to 0.20 $h$Mpc$^{-1}$.  Solid, dashed,
 thick dotted lines indicate 1$\sigma$, 2$\sigma$ and 3$\sigma$
 confidence levels.  Thin dotted lines are the same as those in
 Fig. \ref{fig:simP_constraint}(a).}
\end{figure}

Figure \ref{fig:mock_constraint_k0.07-0.20} plots constraints on the
$(\alpha, \lambda)$ plane derived from the $\Delta \chi^2$ fit to the
SDSS bispectrum using $p^{(3)}$ and assuming a linear bias ($b_2=0$).
The constraints from the bispectrum are fairly consistent with, but
slightly more stringent than, those from the power spectrum, which
indicates the complementary role of the higher-order clustering
statistics.

%%%%%%%%%%%%%%%%%%%%%%%%%%%%%%%%%%%%%%%%%%%%%%%%%%%%%%%%%%%%%%%%%%%%%%%
% Fig. 6
%%%%%%%%%%%%%%%%%%%%%%%%%%%%%%%%%%%%%%%%%%%%%%%%%%%%%%%%%%%%%%%%%%%%%%%
%%%%%%%%%%%%%%%%%%%%%%%%%%%%%%%%%%%%%%%%%%%%%%%%%%%%%%%%%%%%%%%%%%%%%%%

%%%%%%%%%%%%%%%%%%%%%%%%%%%%%%%%%%%%%%%%%%%%%%%%%%%%%%%%%%%%%%%%%%%%%%%
% Fig. 7
%%%%%%%%%%%%%%%%%%%%%%%%%%%%%%%%%%%%%%%%%%%%%%%%%%%%%%%%%%%%%%%%%%%%%%%
\begin{figure}
\begin{tabular}{c}
      \resizebox{80mm}{!}
 {\includegraphics{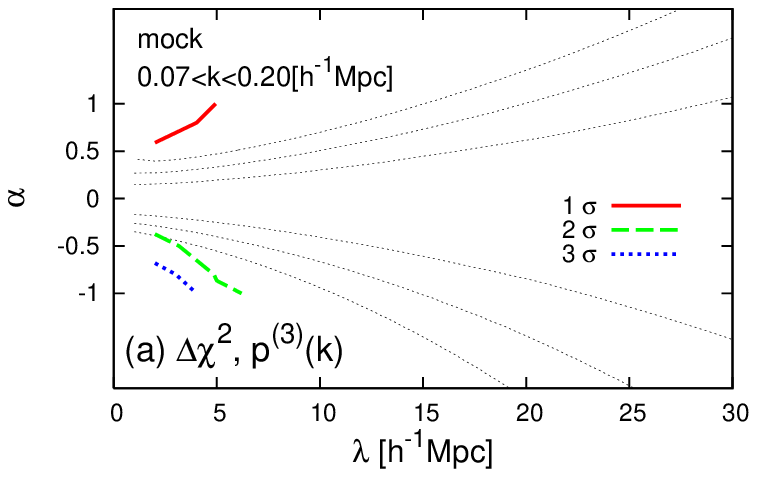}} \\
      \resizebox{80mm}{!}
 {\includegraphics{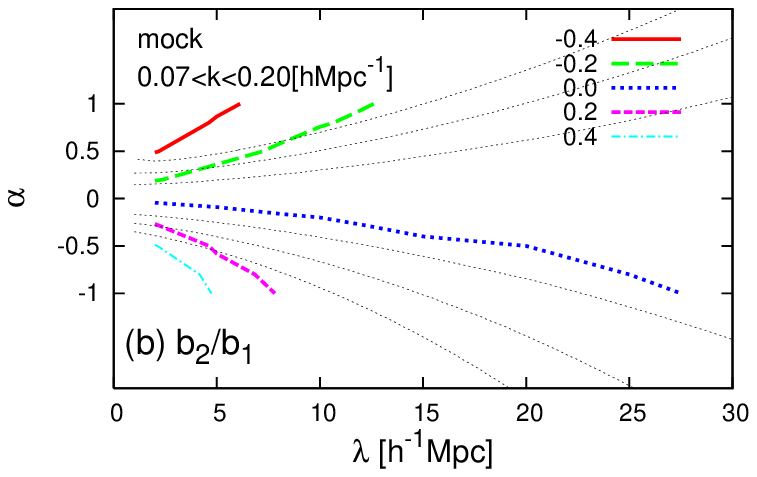}}
\end{tabular}
\caption{ \label{fig:mock_constraint_k0.07-0.20_b2} (a) Constraints on
$\alpha$ and $\lambda$ from the $p^{(3)}$ analysis treating $b_2/b_1$ as
a free parameter. (b) The best fit values
of $b_2/b_1$ that gives minimum $\chi^2$ for $p^{(3)}$.  Thin dotted
lines are the same as those in Fig. \ref{fig:simP_constraint}(a).  }
\end{figure}
%%%%%%%%%%%%%%%%%%%%%%%%%%%%%%%%%%%%%%%%%%%%%%%%%%%%%%%%%%%%%%%%%%%%%%%

\subsection{The effect of non-linear biasing}

In reality, however, it may be more appropriate to analyze the
higher-order clustering statistics adopting a nonlinear bias model.  In
the case of the bispectrum, it implies to introduce the quadratic
biasing parameter $b_2$ [see Eq. \eqref{eq:bias}].
In this bias model, the relation of $p^{(3)}_{\rm g}$ for galaxies and
$p^{(3)}$ for mass reduces to
\begin{gather}
 p^{(3)}_{\rm g} (\bk_1, \bk_2)
= p^{(3)}(\bk_1, \bk_2) + \frac{b_2}{b_1}
f(P_1,P_2,P_3),
\\
f(P_1, P_2, P_3) \equiv \frac{P_1 P_2 + P_2 P_3 + P_3 P_1}
{\sqrt{V_{\rm samp} P_1 P_2 P_3}},
 \label{eq:p3_b2}
\end{gather}
where $P_i = P(k_i)$ for $i = 1,2,3$ \cite{Hikage:2005ia}.

Previous papers \cite{Hikage:2005ia,Nishimichi:2006} suggest that a
simple linear bias model in the Newtonian gravity model describes well
the clustering of the volume-limited sample of SDSS galaxies, i.e., $b_2
\sim 0$ and $b_1 \sim 1$.  We now repeat the similar analysis in the
modified Newtonian model.

Figure \ref{fig:mock_constraint_k0.07-0.20_b2}(a) indicates constraints
on the $(\alpha, \lambda)$ plane by treating $b_2/b_1$ as a free
parameter, which should be compared with Figure
\ref{fig:mock_constraint_k0.07-0.20} for $b_2=0$.  The regions below the
contours are excluded with the corresponding confidence level.
Naturally the bispectrum alone does not constrain $(\alpha, \lambda)$
significantly in this generalized model. While the $\alpha=0$ models are
excluded with a 1$\sigma$ confidence level, the conclusion is not
statistically significant.  In turn, however, we can derive constraints
on the value of $b_2/b_1$ for the modified gravity model by combining
the constraints from power spectrum (independent of the value of
$b_2/b_1$).  Figure \ref{fig:mock_constraint_k0.07-0.20_b2}(b) shows the
contours of the best-fit value of $b_2/b_1$ that gives the minimum
$\chi^2$ for $p^{(3)} (k)$ on the plane.  Figure
\ref{fig:mock_constraint_k0.07-0.20_b2}(b) suggests that $b_2/b_1$
should satisfy $-0.4 < b_2/b_1 < 0.3$, which is the first constraint on
the quadratic biasing parameter in the modified Newtonian model.

\section{Summary}
\label{sec:summary} 

We have derived constraints on possible deviations from Newtonian
gravity using the power spectrum and the bispectrum of Sloan Digital Sky
Survey galaxies.  Our model assumes an additional Yukawa-like term with
two parameters that characterize the amplitude, $\alpha$, and the length
scale, $\lambda$, of the modified gravity.  We have predicted the power
spectrum and the bispectrum using two different methods, the
perturbation theory and direct $N$-body simulations, and
found the good agreement in real space as long as the biasing between
galaxies and mass is neglected. In order to take the biasing effect into
consideration, we adopt a quadratic biasing model.  By comparing with
the mock catalogues constructed from our simulations, we have derived
constraints on $\alpha$ and $\lambda$.  This method allows us to compute
the clustering statistics in redshift space and taking account of
various observational effects such as survey geometry as well.  The
resulting constraints from power spectrum are consistent with those
obtained in our earlier work, indicating the validity of the previous
empirical modeling of gravitational nonlinearity in the modified
Newtonian model. If linear biasing is adopted, the bispectrum of the
SDSS galaxies yields constraints very similar to those from the power
spectrum. If we allow for the nonlinear biasing instead, we find that
the ratio of the quadratic to linear biasing coefficients, $b_2/b_1$,
should satisfy $-0.4 < b_2/b_1<0.3$.  in the modified Newtonian model.

Future observations will exploit large ground-based telescopes to probe
the matter density distribution by weak gravitational lensing. Combined
with data from galaxy redshift surveys, lensing observations will
provide invaluable informations on galaxy bias. Then it will be possible
to put more stringent constraints on deviations from Newton's law of
gravity at cosmological scales, using the methodology presented in the
this paper.

\section*{Acknowledgements}

We would like to thank Atsushi Taruya, Kazuhiro Yahata, Takahiro
Nishimichi, Shun Saito, and Issya Kayo for useful discussions and
comments.  A. S. acknowledge the support from Grants-in-Aid for Japan
Society for the Promotion of Science Fellows.  The simulations were
performed at the Data-Reservoir at the University of Tokyo.  We thank
Mary Inaba and Kei Hiraki at the University of Tokyo for providing the
computational resources.  This work is supported in part by
Grants-in-Aid for Scientific research of the Ministry of Education,
Culture, Sports, Science and Technology (No.~17684008, and 18072002),
and by JSPS (Japan Society for Promotion of Science) Core-to-Core
Program ``International Research Network for Dark Energy''.

\end{document}